# Long-Term Oscillations of Sunspots and a Special Class of Artifacts in SOHO/MDI and SDO/HMI Data

V.I. Efremov[1], A.A. Solov'ev[1,4], L.D. Parfinenko[1], A. Riehokainen[2], E. Kirichek[1], V.V. Smirnova[1,2], Y.N. Varun[4], I. Bakunina[5], I. Zhivanovich[1,3]

[1]Central (Pulkovo) Astronomical Observatory, Russian Academy of Sciences, Russia, St-Petersburg, e-mail: maklimur@gmail.com

[2]University of Turku, FI-20014, Turku, Finland, e-mail: alerie@utu.fi

[3]Sobolev Astronomical Institute, St-Petersburg State University, Russia, St-Petersburg, e-mail: ivanzhiv@live.com

[4]Kalmyk State University, Russia, Elista, e-mail: solov@gao.spb.ru

[5]National Research University Higher School of Economics, Russia, Nizhny Novgorod, e-mail: rinbak@mail.ru

**Abstract** A specific type of artifacts (named as *"p2p"*), that originate due to displacement of the image of a moving object along the digital (pixel) matrix of receiver are analyzed in detail. The criteria of appearance and the influence of these artifacts on the study of long-term oscillations of sunspots are deduced. The obtained criteria suggest us methods for reduction or even elimination of these artifacts. It is shown that the use of integral parameters can be very effective against the *"p2p"* artifact distortions. The simultaneous observations of sunspot magnetic field and ultraviolet intensity of the umbra have given the same periods for the long-term oscillations. In this way the real physical nature of the oscillatory process, which is independent of the artifacts have been confirmed again. A number of examples considered here confirm the dependence between the periods of main mode of the sunspot magnetic field long-term oscillations and its strength. The dependence was derived earlier from both the observations and the theoretical model of the shallow sunspot. The anti-phase behavior of time variations of sunspot umbra area and magnetic field of the sunspot demonstrates that the umbra of sunspot moves in long-term oscillations as a whole: all its points oscillate with the same phase.

Key words Sun, oscillations, sunspots, magnetic field, time series, artifacts

## 1. Introduction

In the last decade we have paid a great attention to the study of long-period oscillations of the magnetic field and the line-of-sight velocities of sunspots (Efremov et al, 2010, 2012, 2014), as well as the temporal variations of emission of radio sources associated with sunspots (Kallunki and Riehokainen, 2012; Smirnova et al, 2011, 2013, Bakunina et al, 2013, 2016). In accordance with the theoretical model of shallow sunspot (Solov'ev and Kirichek, 2014), it was established that sunspots, being long-lived solitary magnetic structures, oscillate as a whole near the position of their stable equilibrium state. These oscillations are excited due to disturbances in the surrounding turbulent environment (the convection zone), and are manifested in the form of long-term quasi-periodic variations of the magnetic field and the line-of-sight velocities registered in the umbra of sunspots. The periods of oscillations at the lowest eigen mode depend specifically on the strength of the magnetic field of the sunspot, and lie in the range from 10-12 hours (when the sunspot field strength is about 2700 Gauss) and up to 32-40 hours (when the field strength is greater or less than 2700 Gauss). Besides the lowest eigen mode, a number of higher harmonics with smaller amplitudes and periods have been observed (Efremov et al, 2012, 2014). Investigation of the physical nature of long-period sunspot oscillations is fundamentally related to the question of whether the umbra of the sunspot oscillates as a whole or in parts. Our preliminary results (Efremov et al, 2012) give the evidence for a synchronous pattern of the sunspot oscillations: the entire umbra of sunspot shifts quasi-periodically along the vertical as a whole. However, this result, because of its fundamental importance, needs further investigation and confirmation.

In the study of long-period oscillations of sunspots, using the data of space based observatories, we are faced with a number of artifacts, some of which are associated with the peculiarities of the orbital motion of the Solar Dynamic Observatory (Smirnova et al. 2013), and others originate due to the peculiarities of registration of the signal from the spatially distributed object moving along the discrete receiver (CCD pixel matrix). The last type of artifacts is caused by the translation of a moving object from one pixel of the matrix to another. Hence we named the phenomenon as *"p2p"* (pixel to pixel) effect (Efremov et al., 2010). This phenomenon has general scientific importance because of its manifestation in any data obtained by digital devices. In particular, it appears in the determination of orbits of comets and meteors (Mechinsky, 2013).

Depending on the direction of motion of the distributed object on the receiver matrix, one should distinguish between *"X-p2p"* and *"Y-p2p"* effects, as the speed of an object (sunspot) along the horizontal $X$ direction on the matrix is much greater than the rate of movement of the source along the vertical $Y$ direction. The movement of the spot along the $Y$ axis of the matrix is due to the slope of the axis of the Sun's rotation relative to the plane of the ecliptic. Consequently, from December $6^{th}$-$7^{th}$ to June $6^{th}$ -$7^{th}$ , a sunspot moves across the solar disk along a path slightly curved up and running above the center of the disk, and from June $6^{th}$ -$7^{th}$ to December $6^{th}$ -$7^{th}$ , the trajectory of a sunspot is curved to the bottom and passes below the center of the disk. Close to $6^{th}$ of December and $6^{th}$ of June sunspots move across the disk parallel to the horizontal axis of the matrix of the receiver, and *"Y-p2p"* effect is absent (further, we will refer to these days as "epoch zero"). Similarly close to March $6^{th}$ and September $6^{th}$, the effect is most pronounced, and we call these days as the "epoch max".

In accordance with the above stated problems, our article considers the following: i) Detailed study of the artifacts (*"p2p"* effects) and their possible impact on the determination of the frequency of long-period oscillations of sunspots; ii) Obtaining additional reliable evidence of the synchronism between umbral oscillations and the magnetic field of the sunspot.

The paper is organized as follows: In Section 2, we give a brief description to construct of a time series from sequence magnetograms (et al.) We define the basic concepts used in the article. In Section 3, the analysis of the physical nature of "p2p" effect is given in detail and basic criteria of its manifestation are derived. Section 4 is especially dedicated to the investigation of slow *"Y-p2p"* artifact and its possible influence on the results of the investigation of long-term oscillations of a sunspot. Section 5 is devoted to the usage of integral parameters of sunspot which allows us to practically eliminate the influence of *"Y-p2p"* artifact. Two additional examples are also provided to prove the synchronous nature of long-period oscillations of the magnetic field of a sunspot inside the "umbra-penumbra" contour. In conclusion, we summarize our results of the study.

## 2. Data Processing Procedures

A method for constructing a time series of physical quantities from the data obtained by SOHO/MDI and SDO/HMI instruments (intensitygrams, magnetograms, ultraviolet data) was described previously in the works: Efremov et al. (2010, 2014, 2016). For example, to form a time series of magnetic-field strength in a sunspot, we used the extreme-value method. The time series of magnetic field are formed by an extreme value of the pixel in the sunspot. These extreme pixels are located according to the topology of a particular sunspot, which is not necessarily located at its geometrical center.

## 3. Artifact "p2p" as a Phenomenon

First of all, it is necessary to discuss the physical reasons for the occurrence of these artifacts. Actually a sunspot is a distributed object whose image can be obtained in any one of the channels of the SOHO or SDO instruments (for example, in a magnetically sensitive line or in an ultraviolet line of the SDO/AIA, 1700 A).

Due to the rotation of the Sun, sunspot moves over the solar disc and along the discrete matrix of the receiver in the *X*- and *Y*-directions. Since the phenomenon has significantly different scales in these two directions, we should distinguish between *"X-p2p"* and *"Y-p2p"* effects. The first artifact *("X-p2p")* mainly is due to the rotation of the Sun and has a great degree of latitudinal dependence, but because of projection effects it has the longitudinal dependence also. The second one (*"Y-p2p"*) is due to an inclination of the axis of rotation of the Sun with respect to the ecliptic plane and has a significant seasonal dependence. Tracks of sunspots on the CCD matrix have a maximum curvature during the days approaching the "epoch max", and parallel to the equator of the Sun and during the days close to "epoch zero", as it was earlier noted in the Introduction.

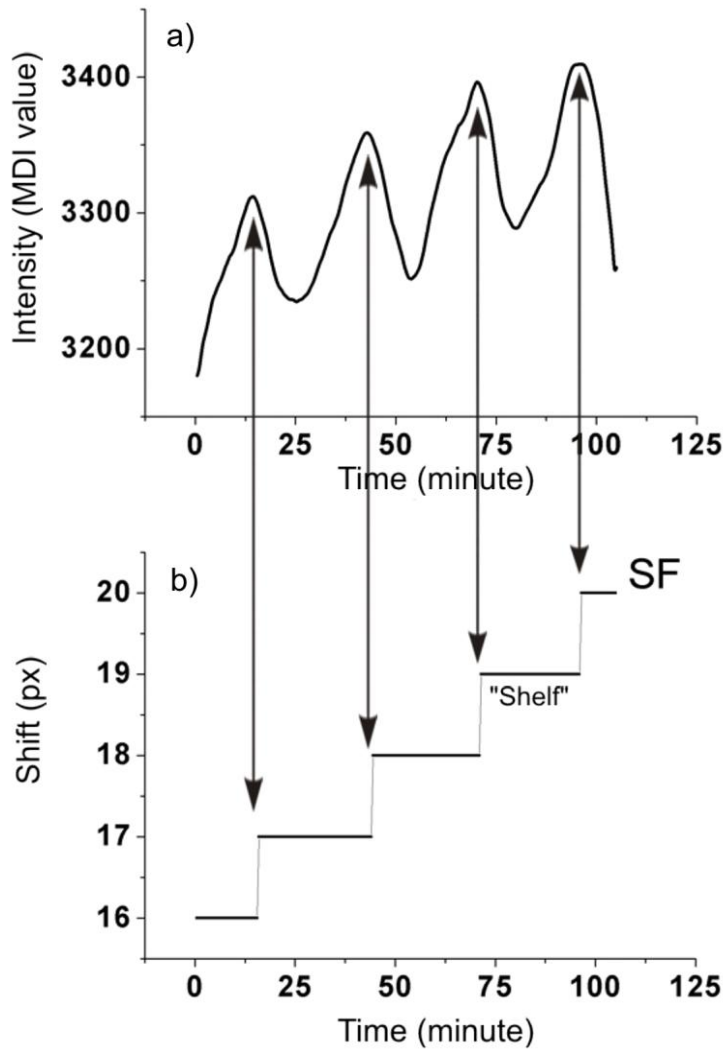

Figure 1 a) Figure shows the changes of intensity in sunspot umbra in the point of the extremal emission; b) Panel one can see a shift function (*SF*) for the point of the extremal intensity moving along the axis *X*. Data for the Figure were obtained with SOHO /MDI instrument in spectral line 6173 A for active region NOAA 08156 on 15 Feb. 1998.

Now we will consider, in a greater detail, the *"X-p2p"* artifact and criteria for its manifestation in the observational data. Figure 1 (top panel) shows the variations of the recorded intensity in the spectral line 6173 A. This record was obtained by SOHO/MDI instrument in *FD-continuous mode* for a point of sunspot umbra which has an extremal intensity (the minimum in this case). In the bottom panel of Figure 1 we show the shift function *SF* for this point as it moves along the *X* axis of receiver's matrix in the vicinity of the central meridian (*CM*). It is seen that the variations of intensity are periodic, and the extreme value of the point at the moment of transition from one pixel to another is somewhat higher than in the central part of the "shelf". This occurs because the pixel receives a signal at the moment of transition from the area surrounding the extremal point where the intensity is slightly higher. When the point of the extreme is located in the center of shelf, we again see the minimum, as it is shown in Figure 1. The duration of the observations was 105 minutes, the image size of the Sun on the CCD matrix was 512 x 512 pixels and the sampling interval for the time series was 30 seconds. The period of intensity variations coincides exactly with the "lifetime" of the extremal point at the given pixel and amounts for 26 minutes. It should be noted that the period of *"X-p2p"* transition depends on the size of the CCD matrix. Hence for a 4096 x 4096 pixel CCD cameras (SDO/HMI) the period near the *CM* will be just 3.6 minutes.

### 3.1. Criteria of the Artifact Manifestations

The experience we have obtained while processing the data from space observatories, *i.e.* SOHO/MDI and SDO/HMI, allows us to derive three qualitative criteria for a clear manifestation of the *"p2p"* artifacts in the signal:

(i)   high cadence in the time-series;
(ii)  short observational internal;
(iii) strong gradient of the field in the plane of the matrix.

### 3.2. Cadence of the Time-series

The discreteness of time series or cadence in an examined time series "*Δt*" is a key factor at the initial stage of the data processing. The very identification of false signal in the studied time-series depends on it. The period T of *"X-p2p"* artifact depends on the rotation speed of the Sun, on the position of the object relative to *CM* and on the size of the matrix of the receiver. When the size of matrix and the position of a sunspot on the solar disc are given, one can uniquely determine the period *"T(X-p2p)"* of artifact. Typically, the transition period is obtained from the observations *i.e.* by processing the time-series. But for some limiting cases, the transition time can be easily calculated. For example, the CCD matrix of the receiver in SOHO /MDI has 1024 x 1024 px. The size of the Sun (the diameter, *Ds*) on such a matrix takes just 960 px ($N_{wr}$) for a given date, for example 18$^{th}$ of March 2001. As the linear velocity $V_{eq}$ of the tracer along the solar equator is 2 km s$^{-1}$, we obtain an estimation for the period of *p2p* transition as $T = Ds/(V_{eq}*N_{wr}) = 1.4*10^6/(120*960) = 12.2$ min. For the SDO matrix, the corresponding value will be 4 times less, i.e. just 3 minutes).

Obviously, if the *Δt* of the investigated time-series is taken less than the value of this period, i.e. if *Δt < T(X-p2p),* the parasitic component appears automatically in the power spectrum of the signal. Geometrically, the *SF* along *X*-axis will take the form of typical "shelves", or "stairs", when the different moments of time correspond to the same values of *X*-coordinate. The ratio of *T(X-p2p)/Δt* determines the height and width of a false peak in the power spectrum. Thus, a high cadence happens to be the necessary condition for the appearance of "*p2p*" artifact.

### 3.3. Observational Interval

The movement of an object along the spherical surface is projected on a flat matrix of receiver, therefore the travel speed and, accordingly, the period of "*X-p2p*" effect depend on the position of the object on the solar disc. Hence the longitudinal rate of change of a "*X-p2p*" transition, $V_x(t)$, is proportional to $cos\theta$, i.e. $V_x(t) = V_0 * cos\theta$, where $V_0$ is the velocity of the object when it passes across the central meridian along the equator. For short durations, ranging from a few to ten transitions, the time period "*T(X-p2p)*" varies very little and we get excess power at the average frequency in the power spectrum of the studied time series. For example, in Figure 2a we show the power spectrum obtained at the fixed the observational interval (15 hours) for different longitudes: 80°, 60°, 30°, 0° (it is *CM*). For these longitudes the periods of "*X-p2p*" are correspondingly: 60, 20, 14, 12.5 minutes. However, if we would take the duration interval from -80° to +80°, (it corresponds to the observational internal of about 10 days) we would obtain a uniform *drift* of frequency on the wavelet map, and in the power spectrum derived by FFT procedure we get a continuous spectrum in the band of frequencies corresponding to the interval of periods from 60 to 12.5 minutes.

In the Figure 2b a test example for a function with a variable period is presented. The wavelet transform and the global spectrum are shown in Figure 2c. This example provides a good mathematical model for the longitudinal dependence of a "*p2p*" effect for the sunspot receding from the *CM*. Let we have an observational interval of length *R* and a small fragment of this time series with the length *r*, so that *r* << *R*. The value *r* may be different at the start and the end of the duration, but it should be sufficiently long to determine the period of the function. Now for this fragment one can define the initial and final drift periods as *T1* and *T2* respectively. The effect of "smearing" or dilation of period "*T(X-p2p)*" in the global wavelet is significant if the following simple relation is valid: *p/q* >> *1*, where *p* = max (*T1,T2*) and *q* = min (*T1,T2*). Indeed, for the process shown in Figure 2b we have taken *p/q* = 120/16 ~ 8, and the condition *p/q* >> *1* is satisfied. As it was expected, the drift of the period is seen in the map of wavelet, and no reliable harmonic can be detected in the global wavelet of the process. Thus, the size of the observational interval plays the role of a smoothing parameter for "*X-p2p*" artifacts (or smearing of it) in the original time-series. Namely, the uniform drift of the parasitic frequency due to the change of object's speed on the receiver matrix is a characteristic feature of the process. It leads to the strong attenuation of its impact on the real components of the time series. Besides, it should be noted that the periods of the real oscillations have no longitude dependence, and this fact greatly distinguishes them from the artifacts.

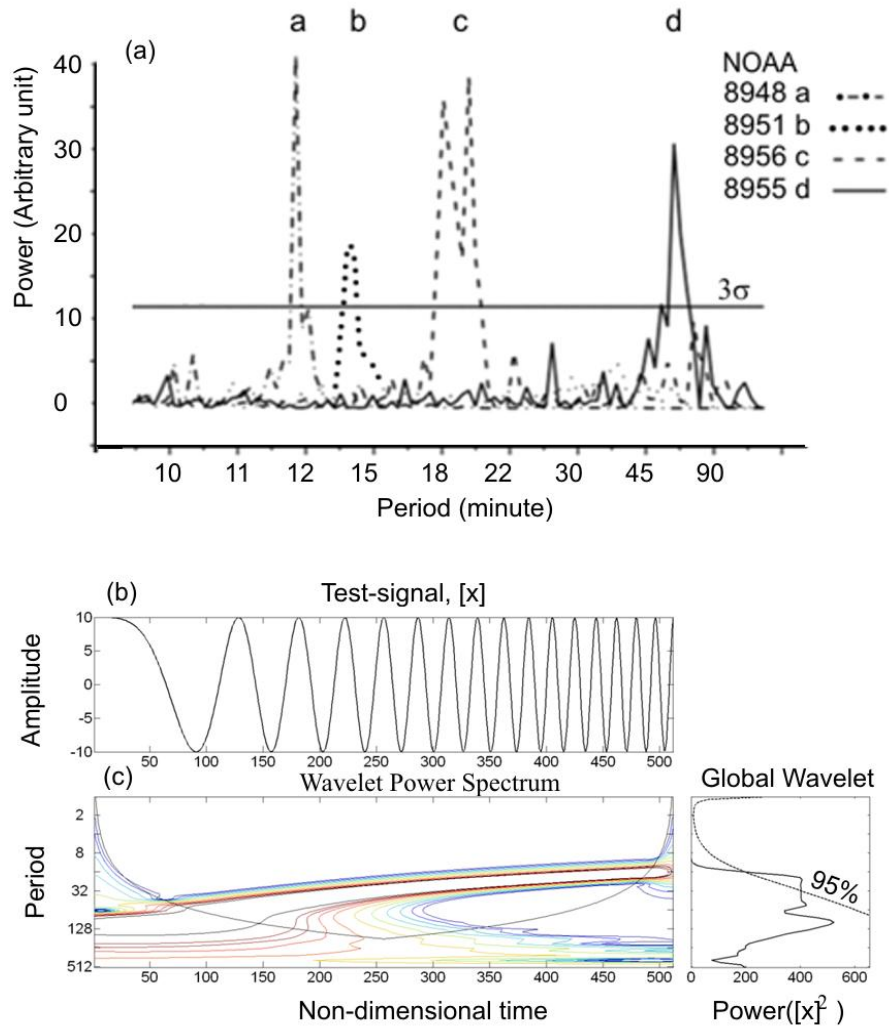

Figure 2 (a) The power spectra for implementation of time series for different longitudes: θ = 80, 60, 30, 0 degrees (periodogram). The periods have values of 60, 20, 14, 12 minutes. (b) The model of the longitudinal dependence for periods *T(X-p2p)* of sunspot, which is moving away from the central meridian (the function with variable period). (c) The figure shows the wavelet transform for this function. On the wavelet map one can see the drift of period, but, on the global panel of the wavelet we do not see any reliable harmonics. A 95% significance level is marked.

### 3.4. Gradient of the Field

As was mentioned above we do not define the field gradient as a vector, but consider it only as a scalar quantity.

Formally, it would seem that when the first of the above necessary conditions $\Delta t < T(X\text{-}p2p)$ is satisfied, the parasitic frequency should appear for any value of the increment. However, it turned out that it is not so because in any real signal a certain level of noise (*W*) is always present. This fact makes some adjustments to the criteria of the occurrence of an artifact. It should be noted that the amplitude itself (or level) of the noise is not important in this case. The ratio of the amplitude of noise (*W*) to the increment of the field ($|\Delta H|$) plays the key role in this case. In this regard, we should introduce one more parameter, so-called "zone of uncertainty" (or *UZ*) defined above as the region of random occurrence of extremal value of the field for a physical value caused by the noise fluctuations. As we shall see below, the size of *UZ*, as well as the size of observational interval, leads to the smearing of *"X-p2p"* artifact. Let us analyze the different relations between ($|\Delta H|$) and *UZ* in terms of artifacts appearance for an object moving on the receiver's matrix. We consider some examples to illustrate the above mentioned relations.

### 3.5. Weak Field Gradient of a Physical Quantity (Flat Bottom)

This case is typical for the data obtained with SDO/AIA UV 1700A. In the central part of the object, the increment of flux $\Delta F(x, y)$ is below than the noise level of the receiver, and it results in the random appearance of an extreme value into some finite region and, consequently, to a smearing of *"X-p2p"* artifact.

For further consideration of the impact of *"p2p"* effects on the spectrum of the signal, it is important to distinguish between the noise of the instrument, *W* (the accuracy of measured values by the instruments) and uncertainty zone *UZ* (see Figure 3, 4a). The first of these is always present while the latter appears only in the case of weak gradients, and does not manifest itself in strong gradients of the field. Figure 3 shows an image of the spot obtained on 18 November 2013 in the Continuum UV 1700A mode with two isophotes taken at the levels of 300 and 400 flux units, and the zone of uncertainty (Grey Square). As noted above, the typical size of *UZ* for these data is about 20x20 pixels, and one can evaluate it in the following way: in figure 4a the function of displacement *X(t)* for *X*-coordinates of an extremal value of the field pertaining to a physical quantity is shown, and as it is seen in the Figure 4a, the scale of the fluctuations is about 20 px ($\delta x_i(t)$). A similar value for the scale of fluctuations will be obtained when we consider the function of displacement (*Y-SF*) along the *Y*-coordinate. Thus, in this case the size of the *UZ* is about 20x20 pixels.

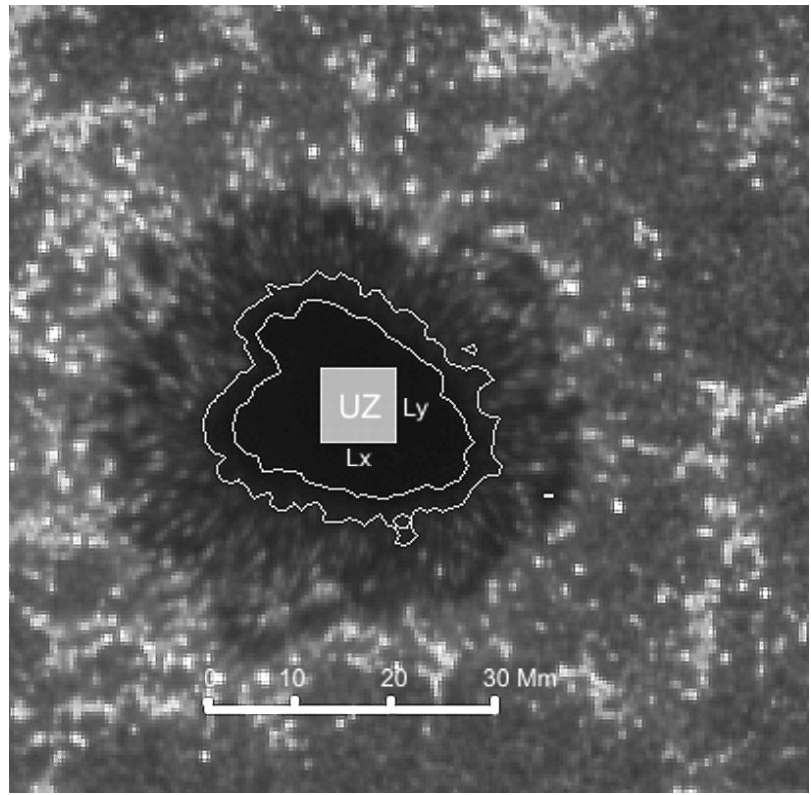

Figure 3  The image of sunspot  (NOAA 11899, 18 November, 2013) obtained in the Continuum UV 1700A with two isophotes at 300 and 400 units of the flux and the *UZ* (grey square) with the size of order  LxLy = 20x20 px.

Despite significant fluctuations of sunspot coordinates, one can clearly notice the trend caused by the motion of a sunspot along the matrix of a receiver. To make it more visible, we used a simple procedure to give a numerical value to the calculated trend in terms of integer representation. This looks like a step function for *"X-p2p"* process (see Figure 4b), *i.e.* resembling the case without any fluctuations. As in this case, the artifact *"X-p2p"* will be manifested for a sufficiently large area of the zone of uncertainty. The form of the step function *X-SF* shows that the size of one shelf of *SF* is equal to 9 points of the time-series. Indeed, in Figure 4b, the spot passes the central meridian. We know that in this case the size of the matrix receiver is 4096x4096 px, and the period of transition *"T(X-p2p)"* is ~ 3.6 minutes. Counting this number in terms of the points of time-series, we obtain: 3.6* 60/24 (Cadence is of 24 sec) = 9 points of time-series. Thus, the trend (Figure 4b), calculated by the shift function *X-SF* (Figure 4a) for *X*-coordinates of an extremal point of a spot demonstrates the *"X-p2p"* transition.

Note again that the position of the extreme point is not defined within the specified square due to the condition  $\Delta F\ (x,\ y) < W$. However, having now the precise function of the displacements along *X*-coordinate, we can calculate the same for the *Y*-coordinate. Forming a pair {X(t),Y(t)} we obtain a scenario of general motion of the spot, *i.e.*, we can now form a signal for the fixed point of the object, in particularly, for any point within the square. It was found, as expected, that these time series are well correlated and they don't differ from each other. It means that, inside the *UZ*, it is not important from what exactly pixel the signal is detected: they are distinguishable only by the amount of noise, by the measurement error of the instrument. However the question of the possibility of spurious signal of type *"T(X-p2p)"* in these time-series remains to be answered.

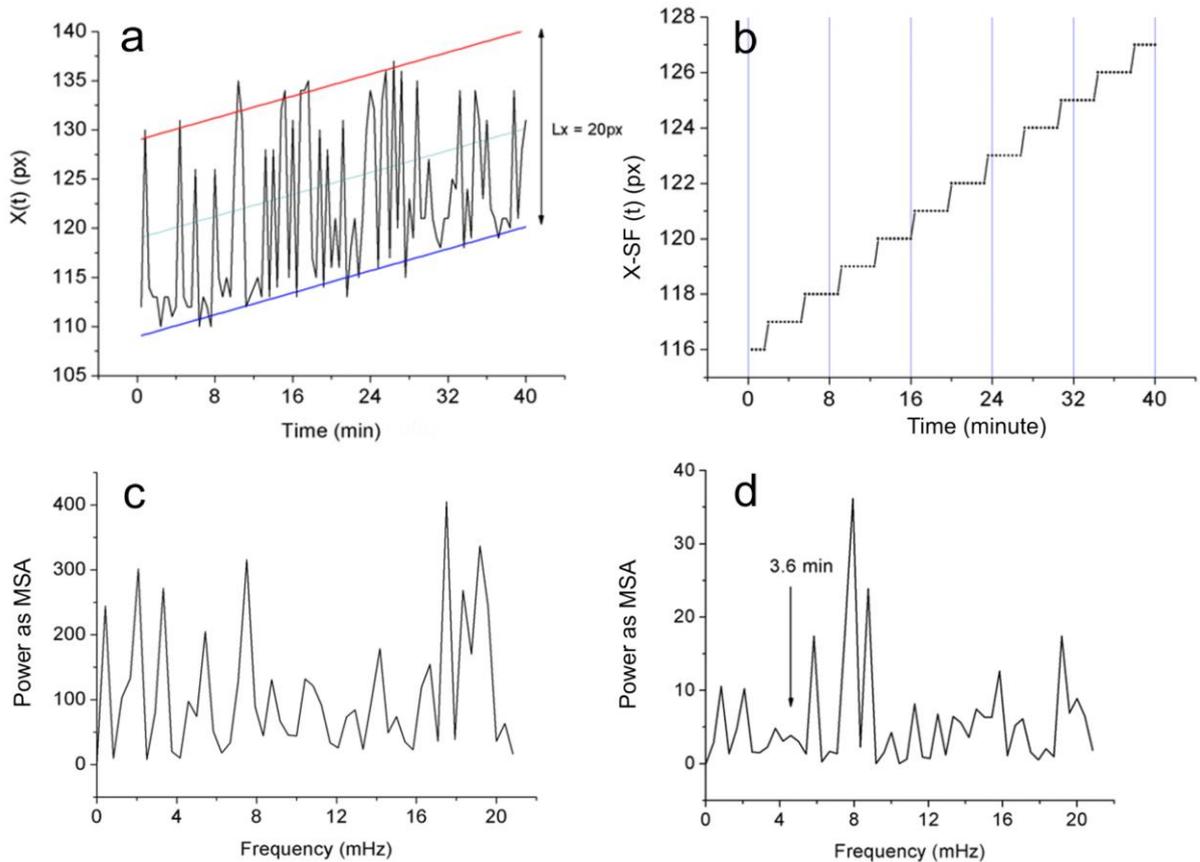

Figure 4 The results of analysis of data, obtained on SDO / AIA UV 1700A (Cadence = 24 sec). a) Behavior of extremum and its trend during the observational interval. b) an integer presentation of the trend. c) Fourier spectrum of the fluctuation part of displacements. d) Fourier spectrum of the real signal in the extremal point.

Figure 4c shows the power spectrum (FFT) of fluctuations of the displacement function *X(t)*. The spectrum does not have any dominant periods and virtually represents a continuous spectrum. This confirms the unpredictable (stochastic) character of occurrence of an extremal value of the field in the region of *UZ*. But the main result is that the power spectrum of the real signal flux *F(t)* (see Figure 4d) has no peak corresponding to the period of 9 points (3.6 min) which is clearly present in the displacement function *X-SF (t)*.

Looking ahead, we note that the random jumping, that is almost always present in the function of *"Y-p2p"* transitions (due to the large duration of this transition), frees the spectrum of a signal from the parasitic component of transitions along the *Y-* coordinate (see Figure 7c, Figure 8a).

It should be noted here that a simple and clear mathematical procedure for the evaluation of the trend in terms of integral representation makes it possible to get the exact form of the step function *X-SF* (Figure 4b, free from noise components and in this way one can find the exact moment of the transition. This procedure will be particularly helpful in the study of *"Y-p2p"* transitions where the transition time of an extremal point of the field is around 5-6 hours.

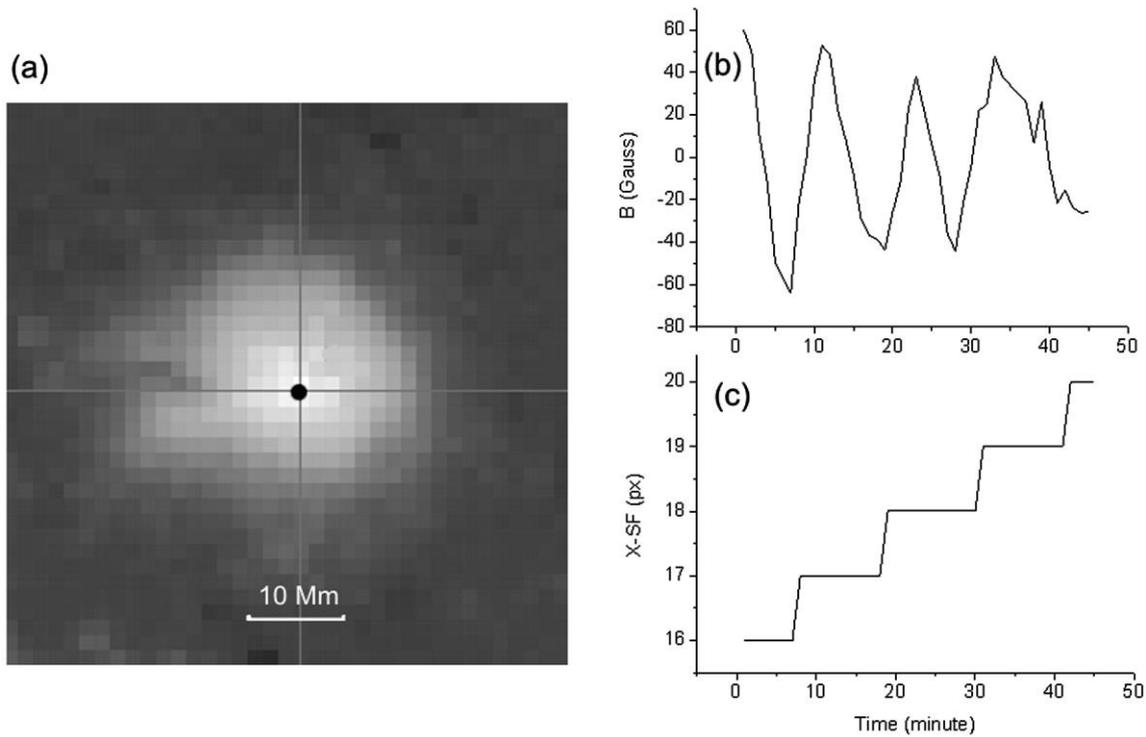

Figure 5  a) A fragment of magnetogram with the sunspot image. The position of the extreme point is marked by the dark dot. b) The change of the magnetic field at the point and c) its coordinate function of displacement.

### 3.6. Strong Field Gradient in SOHO/MDI Data

This situation is typical for the data obtained by SOHO/MDI instrument. The increment of the flux $\Delta F(x, y)$ in neighboring pixels is above the noise level due to relatively low spatial resolution of receiver. This leads to a "hard" localization of the position of extremal values of the field and consequently to the appearance of an *"X-p2p"* effect. We consider, as an example, a short sequence of magnetograms near the *CM* (2002 April 3, NOAA 09887, matrix is 1024x1024 pixels, cadence is 1 m). The spot is located near the equator, and the expected period of transition is: *"T(X-p2p)"* = 13 min. As was mentioned above, it is sufficient to satisfy the inequality: $\Delta t < T(X\text{-}p2p)$ for the appearance of *"p2p"* artifact. The duration of the observations was chosen so that the parasitic periods (artifacts) would be present in the test time series for only a few times so that its period does not change significantly during the observations. Figure 5a shows a magnetogram (*FD_Magnetogram*) of the sunspot in the *CM*. The point with an extremal value of the magnetic field, for which the time series was derived, is marked as the dark dot (Figure 5b). In the same picture (5c) the shift function of *"X-p2p"* transitions is shown. Obviously, the periodical variations of the magnetic field are associated with the transitions of the extremal (maximum) value of the distributed object (sunspot) from one pixel to another, thus at the middle of the shelf, the magnitude of the field is maximum and at the moment of transition it is minimum. As it is expected, the transition period *"T(X-p2p)"* turns out to be about 13.5 minutes.

## 4. Y - p2p Artifact

Now we consider the artifact associated with the transversal displacement of sunspots *i.e.* along the *Y*-coordinate of the matrix receiver. Of course, all aforesaid criteria for the appearance of *"X-p2p"* artifact hold true for the *"Y-p2p"* artifact, with the only difference that the time-scales of the phenomenon change significantly: the parasitic periods *"T(Y-p2p)"* will now be about tens of hours. The fact that the periods of *"Y-p2p"* transitions and the periods of real low-frequency oscillations of the magnetic field in sunspots (Efremov et al. 2012, 2014) are comparable makes the study of long-term oscillatory processes in solar active elements quite difficult. Therefore, it is very important to consider thoroughly what are the specific periods of *"Y-p2p"* artifacts and its strength of impact on the spectrum of a real signal.

### 4.1. Two Limiting States of Y-p2p Artifact Appearance

As we noted above, the reason for the appearance of *"Y-p2p"* artifact is related to the inclination of the solar rotational axis to the ecliptic plane and as a result this phenomenon is strongly dependent on the seasonal observations during the year. Let us designate $B_0$ as the seasonal inclination of poles of the Sun in the plane of the receiver. At a moment approaching the $6^{th}$ -$7^{th}$ of December and $6^{th}$ -$7^{th}$ of June; (we call these periods as "epoch zero") the parameter $B_0 = 0^o$ because of zero inclination. The parameter $B_0$ is named as OBS_B0 and CRLT_OBS for the matrices of the spacecraft SOHO/MDI and SDO/HMI respectively. Due to the orientation of the receiver matrix along the west-east direction, the tracks of the sunspots on the matrix passes parallel to the *X* axis. In the other extremal case of "epoch max" (March $6^{th}$ -$7^{th}$, September $6^{th}$ -$7^{th}$), the axis of rotation of the Sun has a maximum slope of $B_0 = \pm 7^o$ and the tracks of the sunspots have the maximal curvature. Since the projection of sunspot velocity on the matrix will vary uniformly when the sunspot moves along the solar disc, the transition time *"T(Y-p2p)"* will change uniformly. As a result, the "life-time" of the extremal value on the pixel (the width of the shelves) at the moments of the "epoch zero" is equal to the time of observations; and at the moments of the "epoch max" this "life-time" will of course vary. In this case, *"T(Y-p2p)"* will vary in the interval of the maximal value of 500 minutes to 150 minutes (see. Figure 6). For all other sunspot's tracks, the corresponding maximum of *"T(Y-p2p)"* obtained in the observations will be in the interval of [500, *Obs*] minutes, where "*Obs*" is the observational time. A close examination of the test model in Figure 2 can show that there is a small excess of power in the low-frequency region of the spectrum for a period of about 120 minutes in the global wavelet. This value corresponds exactly to the first cycle of the test- function, or, in terms of shift function *Y-SF*, to the maximum width of the "shelf". Although this weak peak in the power spectrum is far beyond the 95% confidence level, some authors (Nagovitsyn, Rybak, 2013), argue that it is a dominant artifact greatly distorting a useful physical signal in the studied time series. Keeping in view the above stated facts we consider two limiting states for the *"Y-p2p"* artifact.

### 4.2. States near the «Epoch Max"

Figure 6 shows a typical displacement of sunspot along the *Y*-axis of the receiver matrix at the time of the "epoch max".

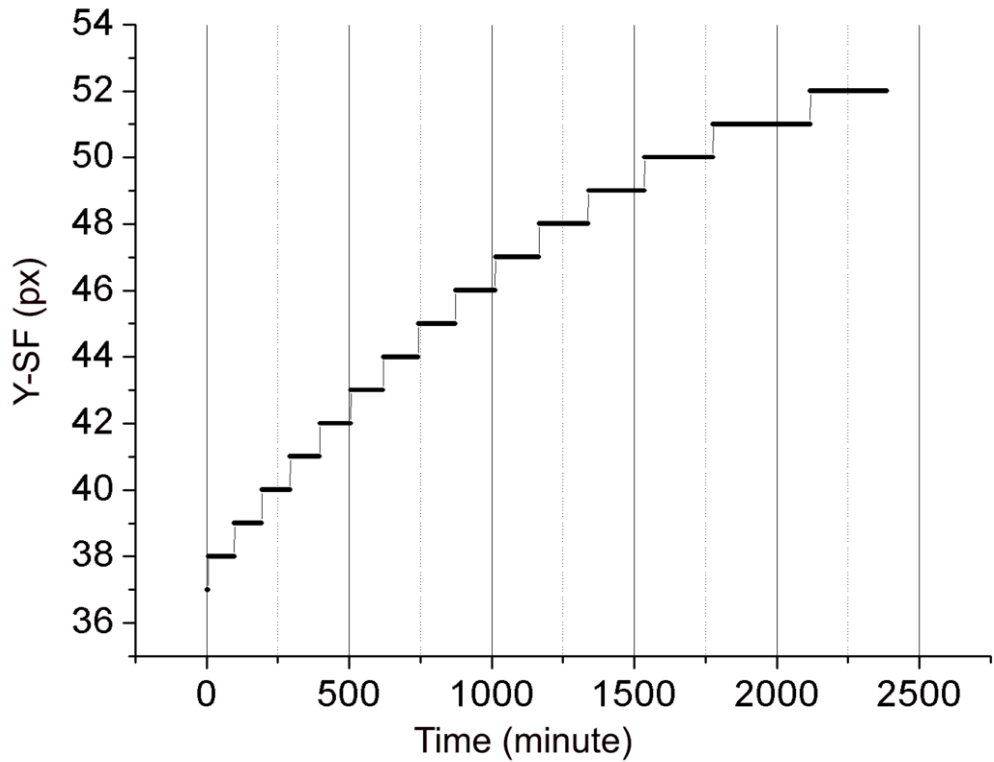

Figure 6 Typical displacement of sunspot along the Y-coordinates of the receiver matrix at the time of the "epoch max".

As the sunspot moves away from the central meridian, the transition time *"T(Y-p2p)"* changes uniformly from a maximum of 500 minutes to a minimum of 100 minutes. This compels us to consider the possibility of a false signal in the spectrum of a time series introduced in the temporal variation of a sunspot's magnetic field with periods close to *"T(Y-p2p)"*. To analyze this, let us consider the sequence of magnetograms received on SOHO/MDI (22-26 September 1999), NOAA 08706, M = 1024x1024 px, Cadence = 1 minute, the duration of observation is *Obs* = 4993 minutes (~ 82 hours) near the "epoch max". Figure 7a shows a time series of the magnetic field strength at the extremal point of the spot (a), its shift function for the *Y*-coordinate *Y-SF(t)* (c), and the corresponding periodograms (b, d). It is clearly seen that the transition time *"T(Y-p2p)"* changes uniformly, but the delay on the *Y*-coordinate during the passage of the spot through the *CM* gives a significant increase of the transit time. In the periodogram (d), we observe a periodic component with the period of 28 hours, corresponding exactly to the size of the maximum "shelf". Finally, the power spectrum is obtained after the removal of the continuous trend. However such an operation in this case is meaningless as the subtraction of the continuous trend from a discrete sequence of integer values *Y-SF* gives a sequence of values of magnitude less than unity. The result (the excess of power in the low-frequency part of the spectrum) so obtained is fully consistent with the above stated remark referring to the test-function of Figure 2b. It is important that there is no period of 28 hours in the time-series of variations of a sunspot's magnetic field. The authors [Nagovitsyn, Rybak, 2013] argue that each transition of a function of *"Y-p2p"* results in a peak in the power spectrum, and the low-frequency peak is the most pronounced. In their view, the *"Y-p2p"* artifact must necessarily be present in the power spectrum of a sunspot's magnetic field variation. But we see that their statement is false because the dominant period of the magnetic field in this case is about 14 hours (Figure 7b). It is just the same value of the period for a dominant mode of the sunspot oscillations with the magnetic field strength of about *B = 2800 G*, that can be expected in accordance from the theoretical model of "shallow sunspot" (Solov'ev & Kirichek, 2014) and from the observational dependence of the periods of long-term sunspot oscillations on the strength of its magnetic field (Efremov et al. 2014). Thus, in the time series derived from the observations near the "epoch max", there is no dominant parasitic signal of *"Y-p2p"* type. Also, this is due to the fact that we are dealing here with the drift of frequency which is one of the factors resulting in the blurring of *"Y-p2p"* artifact (as it was discussed above).

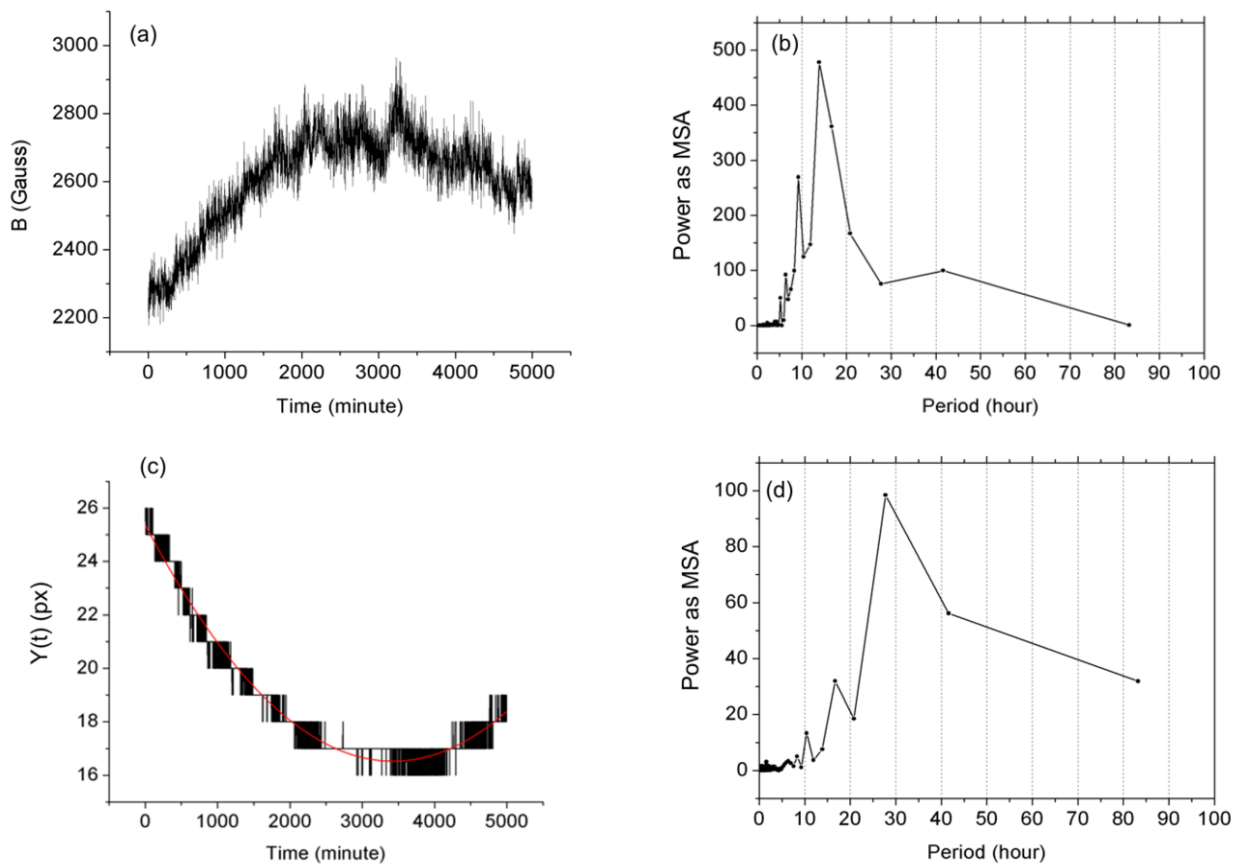

Figure 7  (a) The time-variations of magnetic field for the sunspot NOAA 08706 (22-26 September 1999), and (b) periodogram of these variations derived after the removing the trend. (c) Y(t) function and (d) its periodogram.

## 4.3. States near the "Epoch Zero"

Figure 8a shows a typical shift of the *Y*-coordinates of a spot on the matrix of a receiver at the moments near the "epoch zero". The time-series of the sunspot's magnetic field were obtained with SOHO/MDI, 2-7 December 2002, for NOAA 10209, M = 1024x1024 px, Cadence = 96 minutes, the duration of the observations is *Obs* = 128 hours (Figure 8c). The track of the sunspot is almost parallel to the solar equator, and as it is shown in Figure 8a, the *Y*-coordinate changes slightly. The size of the "shelf" is equal to 52 hours. It is clearly seen on the wavelet transform (Figure 8b) when we repeat again the same doubtful operation to remove a continuous trend from the discrete *Y-SF* function.

However, there is no period of 52 hours in the spectrum of the magnetic field of the time series. As it is shown Figure 8c, the power spectrum of the magnetic field variations has a dominant period of 40 hours! Thus, the time-series derived from the observations made near the "epoch zero" have no dominant parasitic signal of *"Y-p2p"* type. In accordance with the results of (Efremov et al. 2014), the period of 40 hours should be associated with the *M2* mode of sunspot oscillations, which is not an eigen mode of the sunspot, but happens to be a specific mode induced by the perturbations of a sunspot caused by the motions of supergranulation cells surrounding it. However, there is another peak in the power spectrum of the magnetic field which is absent in the function of the *Y-SF*. This peak corresponds to a period of 12 hours. Namely the period of this eigen oscillatory mode should be expected for the sunspot with magnetic field strength of *B = 2650 G*, in accordance with the model of "shallow sunspot" (Solov'ev & Kirichek, 2014) and from the empirical dependence $T = f(B)$ obtained in (Efremov et al.2014).

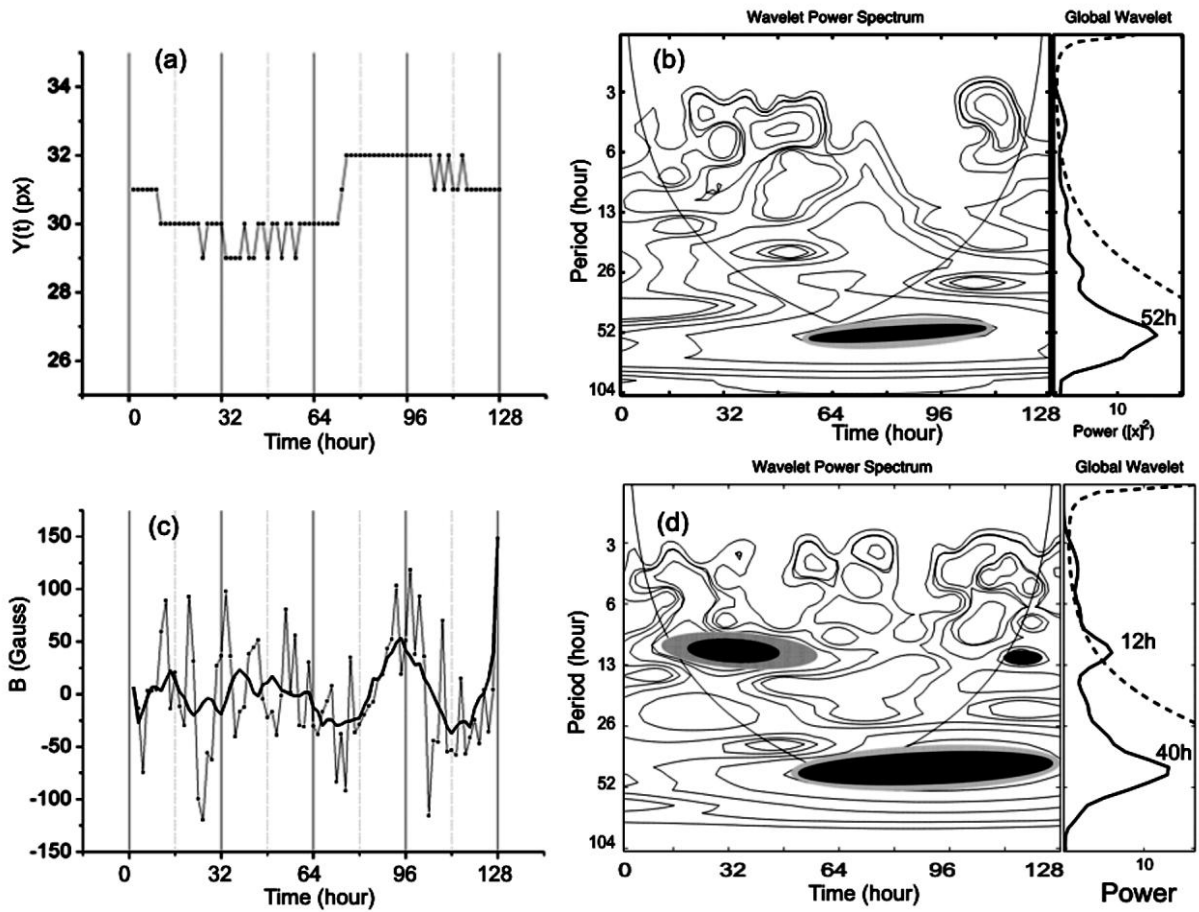

Figure 8 (a) Typical shift of Y-coordinates of the spot on the matrix of the receiver at the time near the "epoch zero" and its periodogram (b). (c) Time series of the magnetic feld of the sunspots derived for the sequence of magnetograms obtained by SOHO/MDI on 2-7 Desember 2002 for active region NOAA 10209, Matrix size= 1024x1024 px, Cadence = 96 minutes, the observational interval Obs = 128 hours, near the "epoch zero" and (d) its periodogram.

## 5. Integral Parameters

Above, we carefully studied and described the criteria for the appearance of *"XY-p2p"* artifacts in the time series. We demonstrated that in some cases, the artifact does not manifest itself, in some cases it is strongly weakened, but there are the cases when it manifests itself significantly. Now the question that obviously arises is regarding the neutralization of the impact of these artifacts on the observed parameters. We will try to find the answer to this question by considering the integral parameters of sunspots, such as the area, the average magnetic field, etc.

## 5.1. Anti-Correlation of the Time-Series at the Front and Backfront Points of the Fixed Contours

We consider again an example to demonstrate the role of integral parameters. We take the time sequence of intensitygrams (16 February 1998, NOAA 08156, M(Image size) = 512x512 px, Cadence = 30 sec).

The expected value of the period of transition *"T (X-p2p)"* near the *CM* is about 25 minutes. As was noted above, the magnitude of the cadence does not affect the value of *"T(X-p2p)"*. The appearance of an artifact is mainly dependent on the condition $\Delta t < T(X-p2p)$, but the size of the matrix affects the transition time because this time depends on the number of pixels. The duration of observations was chosen so that the period of the artifact will be revealed up in the test time series for only a few number of times, so that the parasitic period would not be changed significantly during the observations.

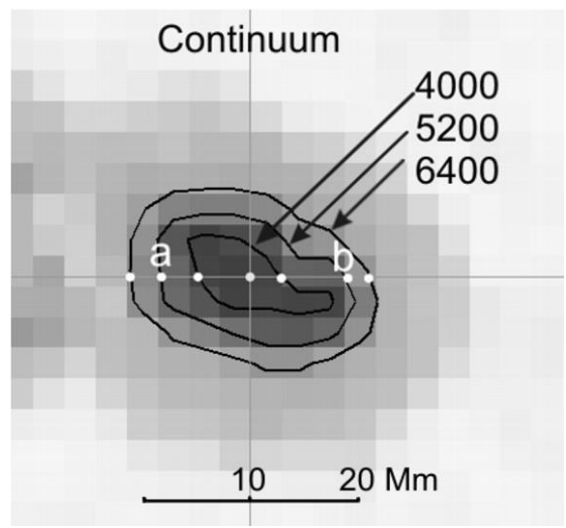

Figure 9  The images of intensity of the sunspot (16/02/1998, NOAA 08156).The contours show the isophotos of intensity (in arbitrary intensity units).The points where the time series were built are marked as white dots.

Figure 9 shows the intensity images of a spots (obtained in *FD Continuum* mode). The contours show the set of isophotos for the intensity values in the spot (aiu; in *Arbitrary intensity units*), the points on the images show places for which the time series were derived. Assuming that the spot moves from right to left along the *X*-axis, let us choose a pair of points (a, b), belonging to the same isophote and lying on opposite sides of the axis *Y* (white dots in Figure 9) and then construct the time series for these points. Let the point "a" (front) is on the left, *i.e.,* at the front, and the point "b" (backfront) is on the opposite side. Figure 10 shows the curves "a" and "b" giving the variations of the intensity of a pair of points of the isophote 5200 aiu , their average (curve "c"), and the shift function *SF*. Let us describe the transition process for the point "a". The digital transition to the neighboring pixel in the direction of movement leads to a sharp increase of the field's value in the pixel as we are moving to higher values of the field, and further, it drops uniformly asymptotically to a value of 5200 aiu, corresponding to the given isophote. The reverse process takes place at the point "b", when it is shifted to the left, so we find ourselves in the region of a weak field, which then converges uniformly and asymptotically to the isophote of 5200 aiu (curve "b"). The process is then repeated, and the regular artifact component appears in the spectral composition of the signal.

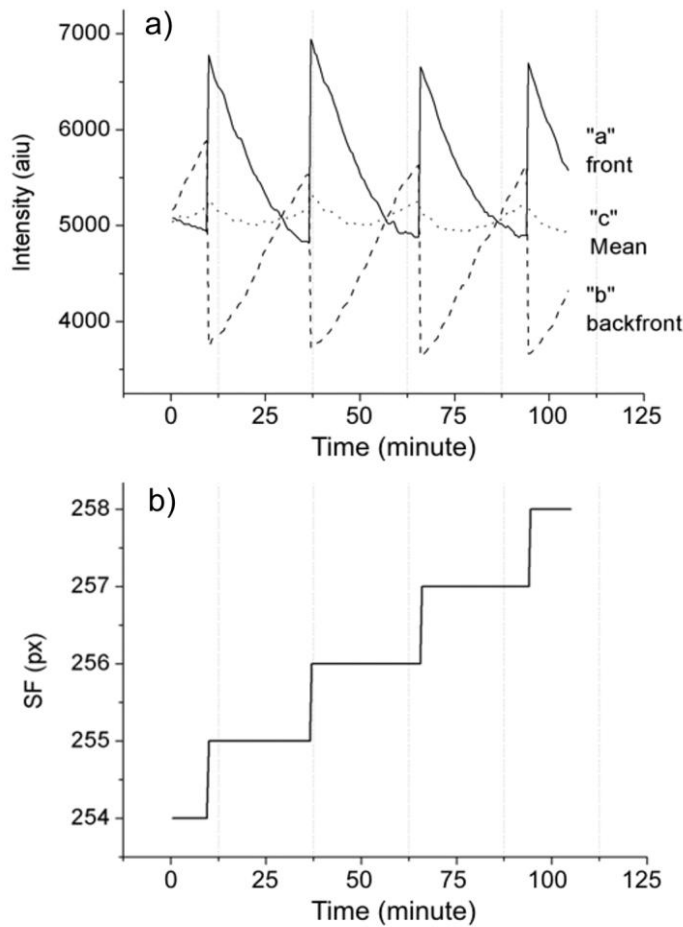

Figure 10 a) The time series of intensities at the points «a» and «b» of the isophote of 5200 aiu, and b) their shift function *SF*.

As was noted earlier, the increment of the field $\Delta F(x, y)$ in the pixel structure of the object plays a significant role in the formation of an artifact. So, if the amplitude of intensity that changes in the point of extremal count is of 200 aiu, then this drop in the considered points ("a" and "b") will be around 2500 aiu (Figure 10a). Such high values of the amplitude, even due to small changes of the field at the points belonging to the same isophote, lead to an incomplete compensation of the parasitic effects (curve for the "c" point in Figure 10a). Despite this, the anti-correlation of the considered curves (for the points "a" and "b") is quite high: *R* (Pearson coefficient of correlation) = - *0.93*. (Indeed, it is difficult to expect a high symmetry for a real sunspot). However, we note that the amplitude of variations for the compensative curve ("c") is only 200 aiu, which is almost an order of magnitude smaller than at the points "a" and "b". In the case of a perfectly symmetrical object (with respect to the axis *Y*), we would have the same increment of field at the points "a" and "b" and hence the full compensation of the parasitic signal. The similar effect of artifact compensation will take place for all other pairs of points belonging to any other isophote.

Thus, with respect to the direction of movement of the object, one can find sunsets of the isophote in which the elements have a pair of conjugated points, whose time series is highly anti-correlated. The artifact will be greatly weakened in the time series, obtained as a sum of the series in these conjugated pairs. In case of a perfectly symmetrical object, the artifact will be fully suppressed. This means that even in the case of a strong gradient of the field of the object (as it could be observed in sunspots at low space resolutions), one can either suppress or even completely eliminate the *"p2p"* artifact from the time series obtained for the integral parameters, such as the area, the average magnetic field, by the summation carried out over entire umbra of the sunspot or over a large part of the umbra.

## 6. The Long-Period Oscillations of Sunspot Parameters

### 6.1. The Antiphase Oscillations of Magnetic Field and the Area of Sunspot Umbra

Now we consider the time-variations of an integral sunspot parameter such as the area of the sunspot umbra. We will use the time sequence of the magnetograms obtained on the SOHO/MDI during 26-29 of June 2000 for the active region NOAA 09056 (the matrix size is of M = 1024x1024 pixels, the Cadence is 1 minute). As in the previous cases, we built the time series for the magnetic field variations in the point of extremal count and for the change of area of the sunspot umbra *i.e.* the area inside the contour "umbra-penumbra". In the figure 8a, we have shown by double arrows the variation of the corresponding values in the anti-phase state. The cross-correlation (*XWT*, "b") and the coherent (*WTC*, "c") wavelet diagrams are presented in the figure 11 (Grinsted, et al, 2004). It is seen that the power of processes is concentrated mainly near 1000 minutes (or 17 hours). It is important to emphasize that the oscillatory processes for the area and for the magnetic fields are clearly anti-correlated.

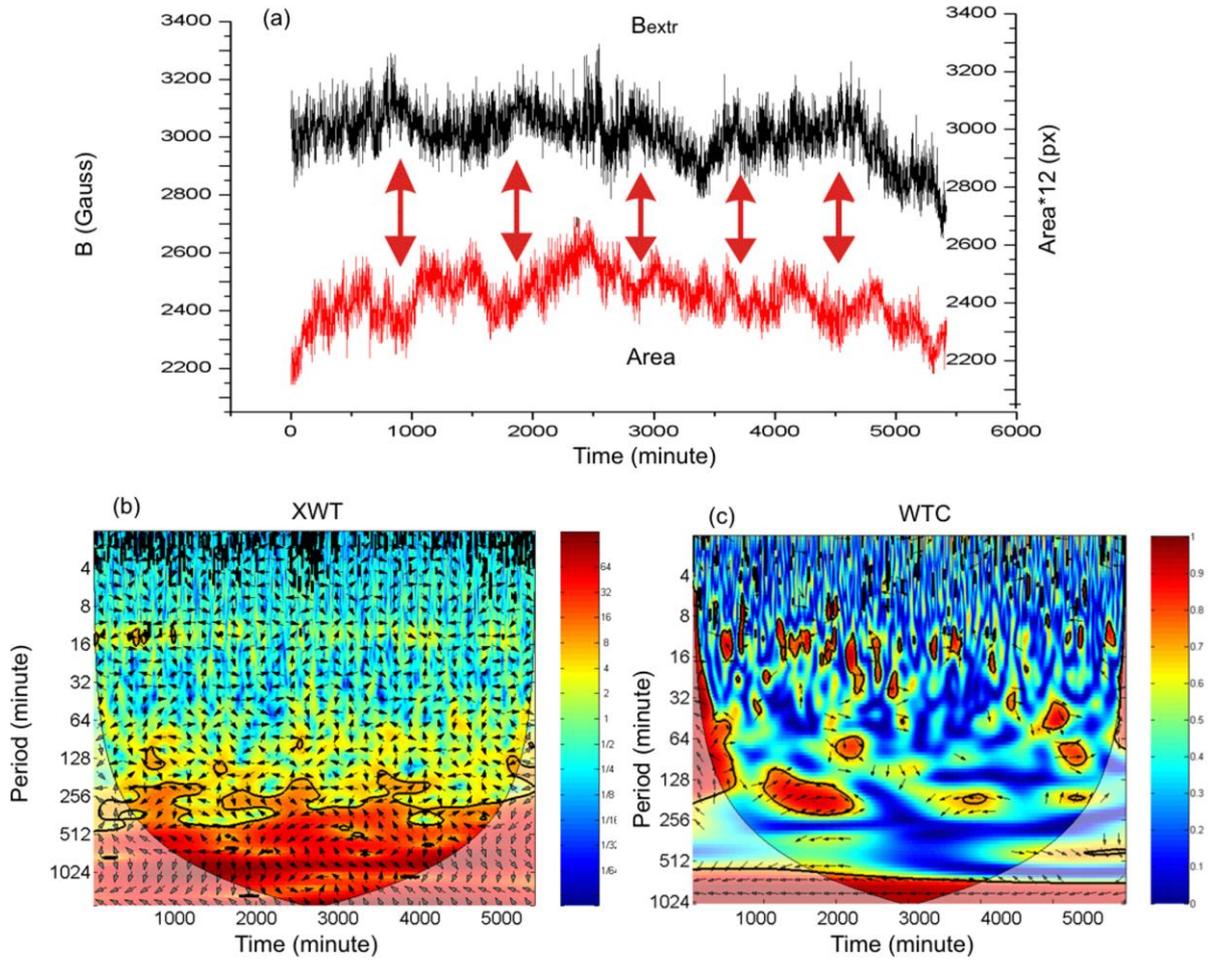

Figure 11 (a) Time series for the magnetic field and the area of umbra of sunspot in active region NOAA 09056 obtained simultaneously on the SOHO/MDI on 26-29 June 2000. (b) Crosswavelet (XWT) and (c) Wavelet Coherence (WTC) for these time series.

According to concept of magneto-gravitational oscillations of a sunspot (the model of shallow sunspot, Solov'ev & Kirichek, (2014), the main oscillatory mode consists of periodical displacements of sunspot umbra as a whole along the vertical. During such oscillatory processes due to the conservation of the magnetic flux inside the contour "umbra-penumbra", the time-variations of the averaged umbral area and averaged magnetic field strength of the sunspot should be in anti-phase with respect to each other. Just this phenomenon is observed in the present case.

## 6.2. Synchronous Observations of Magnetic Field and Ultraviolet Image of a Sunspot

According to the criteria for the occurrence of an *"X-p2p"* artifact, the ultraviolet observations in the line 1700A are unaffected by this artifact because of the high frequency range of the spectrum which results in a large zone of uncertainty zone (*UZ*). The same conclusion applies to the low frequency part of the spectrum, where the appearance of the *"Y-p2p"* artifact could be expected but it is absent because the changes of *H(x,y)* are very weak, $|\Delta H| < W$, and *UZ* is sufficiently large. This favorable circumstance gives us the possibility to compare the spectral components of time series obtained simultaneously for the ultraviolet and for the magnetic field data. We consider the data obtained from SDO/HMI and SDO/AIA on 05 June 2011 for the active region NOAA 11232. The size of the matrix is M=4096x4096 pixels for both the receivers but the cadence is 1 minute for SDO/HMI and 30 minutes for SDO/AIA. The time series and the wavelet transform for these two simultaneous observations are presented in Figure 12. As we see, the dominant periods of these oscillations coincide, and are equal to 34 hours. Namely this value of the period for lower mode of eigen oscillations of sunspot magnetic field should be expected theoretically (Solov'ev & Kirichek, 2014) and empirically (Efremov et al, 2014), when its strength is around *B = 1850* Gauss. In no case this period could be originated due to *"Y-p2p"* artifact, because it is simultaneously present both in the magnetic field, and in the ultraviolet data which are free of artifact.

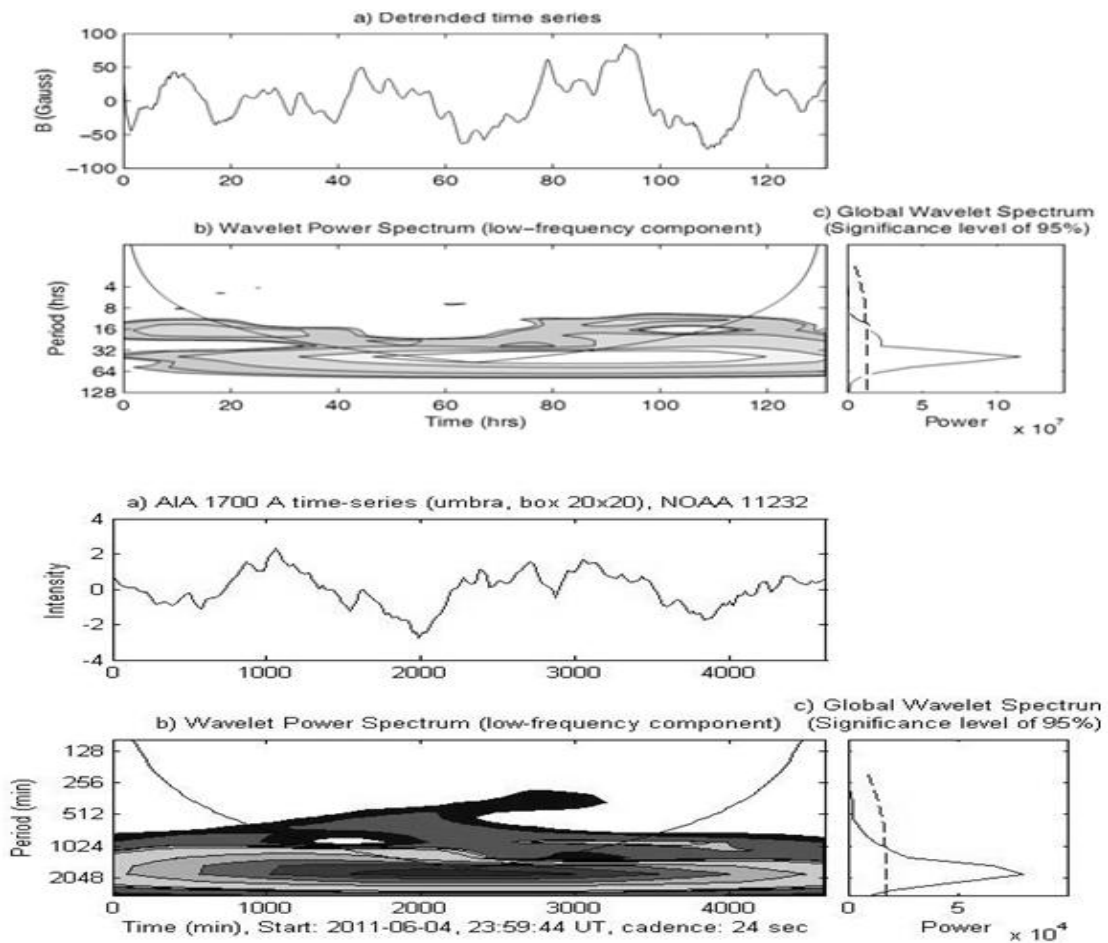

Figure 12 The time series and the wavelet transform for two simultaneous observations data obtained from SDO/HMI and SDO/AIA on 05 June 2011 for the active region NOAA 11232

## 7. Conclusions

The detailed analysis of the *"p2p"* artifacts caused by the displacement of an object along *X*- and *Y*- axis of the digital matrix showed that they can actually take place in observational data, and can affect the results of studies of low-frequency (or long- period) oscillations of sunspots. The criteria for the manifestation of these artifacts have been formulated. We distinguish the two types of transitions resulting in the artifact appearance: the *"X-p2p"* and *"Y-p2p"* transitions.

1. *"X-p2p"*. This artifact has an impact on the relatively high frequency part of the spectrum of oscillations and it has no effect on the low frequency part of the spectrum of sunspot oscillations if we deal with long observational internal (the time series of 5-6 days). For the high-frequency part of the spectrum of the signal this artifact may have a significant impact because in this case the periods of processes under consideration (3-5 minutes) are in the region of the manifestation of this artifact. It should also be noted that these artifacts have a strictly defined period for the short time-series; therefore they can be easily identified and taken into account.

2. *"Y-p2p"*. This artifact corresponds to the low-frequency part of the spectrum of sunspot's signal, therefore long periodical oscillations of sunspot parameters may be subject to the artifact influence. However, we have derived the empirical criteria for the manifestation and the influence of this artifact on the real signal, and have suggested the methods for the significant reduction of the effect due to this artifact.

3. When we consider the integral parameters of the signal (such as the averaged magnetic field inside the contour of a fixed field strength, the area of sunspot umbra, the intensity averaged over some area and etc.), the artifacts of both types turn to be significantly weakened or even completely eliminated. Thus, we believe that the use of the integral parameters allows us to obtain the periods of sunspot oscillations free from *"p2p"* artifacts.

4. The simultaneous observations of sunspot magnetic field and ultraviolet intensity of its umbra gave the same periods of the long-term oscillations. It confirms again the real physical nature of the oscillatory process, independent of the artifacts.

5. In some cases, when we would like to reveal the oscillations of sunspots and nearby structures, we use the data of devices in which the method of data obtaining is significantly different from the usage of discrete CCD matrix. For example, the radio observations which are in principle free of *"XY-p2p"* artifacts (Smirnova et al. 2011, 2013) give the additional evidence of the real physical nature of obtained periods of a signal.

6. A number of examples considered in the paper confirm the dependence of the periods of main mode of long-term oscillations of sunspot on the strength of its magnetic field, as derived earlier from the observations (Efremov et al, 2014) and from the theoretical model of "shallow sunspot" (Solov'ev & Kirichek, 2014).

7. The anti-phase behavior of long-term oscillations of the sunspot area and the averaged magnetic field proves again the integral nature of the oscillations, when different points of the sunspot umbra oscillate synchronously (Efremov et al, 2012).

### Acknowledgements

Authors thank the teams of SOHO and SDO projects for the opportunity to use the data of these space observatories. The work was supported by the Program P-7A and S.School -7241.2016.2.

A. Solov'ev and V. Smirnova thank the Russian Scientific Foundation for the support of the study (projects № 15-12-20001 and 16-12-10448)

Disclosure of Potential Conflicts of Interest The authors declare that they have no conflicts of interest

**Referenses**

Bakunina, I. A., Abramov-Maximov, V. E., Nakariakov, V. M., Lesovoy, S. V, Solov'ev, A. A., Tichomirov, Yu. V., Melnikov, V. F., Shibasaki, S., Nagovitsyn, Yu. A., Averina, E. L.: 2013, Long-Period Oscillations of Sunspots by NoRH and SSRT observations, *Publ. Astron. Soc. Japan.*, 65 No. SP1, Article No.S13, 12. doi: 10.1093/pasj/65.sp1.S13

Bakunina, I. A., Abramov-Maximov, V. E. and Smirnova, V. V.: 2016, The long-term oscillations in sunspots and related inter-sunspot sources in microwave emission. International Conference on Particle Physics and Astrophysics (ICPPA-2015). *Journal of Physics* Conference Series, 675 032029 doi:10.1088/1742-6596/675/3/032029

Efremov, V. I., Parfinenko, L. D., Solov'ev, A. A.,: 2010, Investigation of long-period oscillations of sunspots using ground-based observations (Pulkovo) and instrumental MDI (SOHO) data. *Solar Phys.*, 267, 279. doi: 10.1007/s11207-010-9651-z

Efremov, V. I., Parfinenko, L, D., Solov'ev, A. A.: 2012,Synchronism of long-period oscillations in sunspots. *Geomagnetism and Aeronomy*, 52, 1055. doi: 10.1134/S0016793212080087

Efremov, V. I., Parfinenko, L. D., Solov'ev, A. A., Kirichek, E.: 2014, Long-Period Oscillations of Sunspots Observed by SOHO/MDI. *Solar Phys.*, 289, 1983. doi: 10.1007/s11207-013-0451-0

Efremov, V.I., Parfinenko, L.D., Solov'ev, A.A.: 2016, Ultra Low-Frequency Oscillations of a Solar Filament Observed by the GONG Network, *Solar Phys.* 291, 3357;

Grinsted, A., Moore, J. C., Jevrejeva, S.: 2004, Application of the cross wavelet transform and wavelet coherence to geophysical time series, *Nonlin. Process. Geophys.*, 11, 561566

Kallunki, J., Riehokainen, A.: 2012, *Solar Phys.*, 280, 347. doi: 10.1007/s11207-012-0021-x

Mechinsky, V.: 2013, Satellite photometry. *http://soft.belastro.net/files/sat/satellite photometry.pdf*

Nagovitsyn, Yu. A., Rybak, A. L.: 2014, The properties of long-term sunspot oscillations. *Astronomy Reports*, **58**, 328. doi: 10.1134/S1063772914050047

Otsu, N.: 1979, A threshold selection method from gray-level histograms. *IEEE Trans. Syst. Man Cybern*. 9,62. DOI.

Torrence, C., & Compo, G. P.: 1998, *Bull. Am. Meteorol. Soc*., 79, 61

Smirnova, V. V., Riehokainen, A., Ryzhov, V., Zhiltsov, A., Kallunki, J.: 2011, *Astron. Astrophys.*, 534, id.A137, 6. doi: 10.1051/0004-6361/201117483

Smirnova, V. V., Riehokainen, A., Solov'ev, A. A., Kallunki, J. and Zhiltsov, A.: 2013, Long quasi-periodic oscillations of sunspots and nearby magnetic structures. *Astron. Astrophys.*, 552, A 23S. doi: 10.1051/0004-6361/201219600

Solov'ev, A. A., Kirichek, E. A.: 2014, Basic properties of sunspots: equilibrium, stability and eigen oscillations. *Astrophys. and Space Science*, 352. No.1, 23. doi: 10.1007/s10509-014-1881-3